\documentclass[12pt]{article}

\usepackage{times}
\usepackage{graphicx}
\usepackage{amssymb}
\usepackage{amsthm}
\usepackage{amsmath}
\usepackage{dsfont}
\usepackage{bm}
\usepackage{mathrsfs}
\usepackage{bbold}

\def\XXint#1#2#3{{\setbox0=\hbox{$#1{#2#3}{\int}$ }
\vcenter{\hbox{$#2#3$ }}\kern-.6\wd0}}

\begin{document}

\title{The case for a Casimir cosmology}
\author{Ulf Leonhardt\\
Department of Physics of Complex Systems,\\
Weizmann Institute of Science,\\ Rehovot 7610001, Israel
}
\date{\today}
\maketitle

\begin{abstract}
The cosmological constant, also known as dark energy, was believed to be caused by vacuum fluctuations, but naive calculations give results in stark disagreement with fact. In the Casimir effect, vacuum fluctuations cause forces in dielectric media, which is very well described by Lifshitz theory. Recently, using the analogy between geometries and media, a cosmological constant of the correct order of magnitude was calculated with Lifshitz theory [U. Leonhardt, Ann. Phys. (New York)  {\bf 411}, 167973 (2019)]. This paper discusses the empirical evidence and the ideas behind the Lifshitz theory of the cosmological constant without requiring prior knowledge of cosmology and quantum field theory.  
\end{abstract}

\newpage

\section{The problem}

\begin{figure}[h]
\begin{center}
\includegraphics[width=32pc]{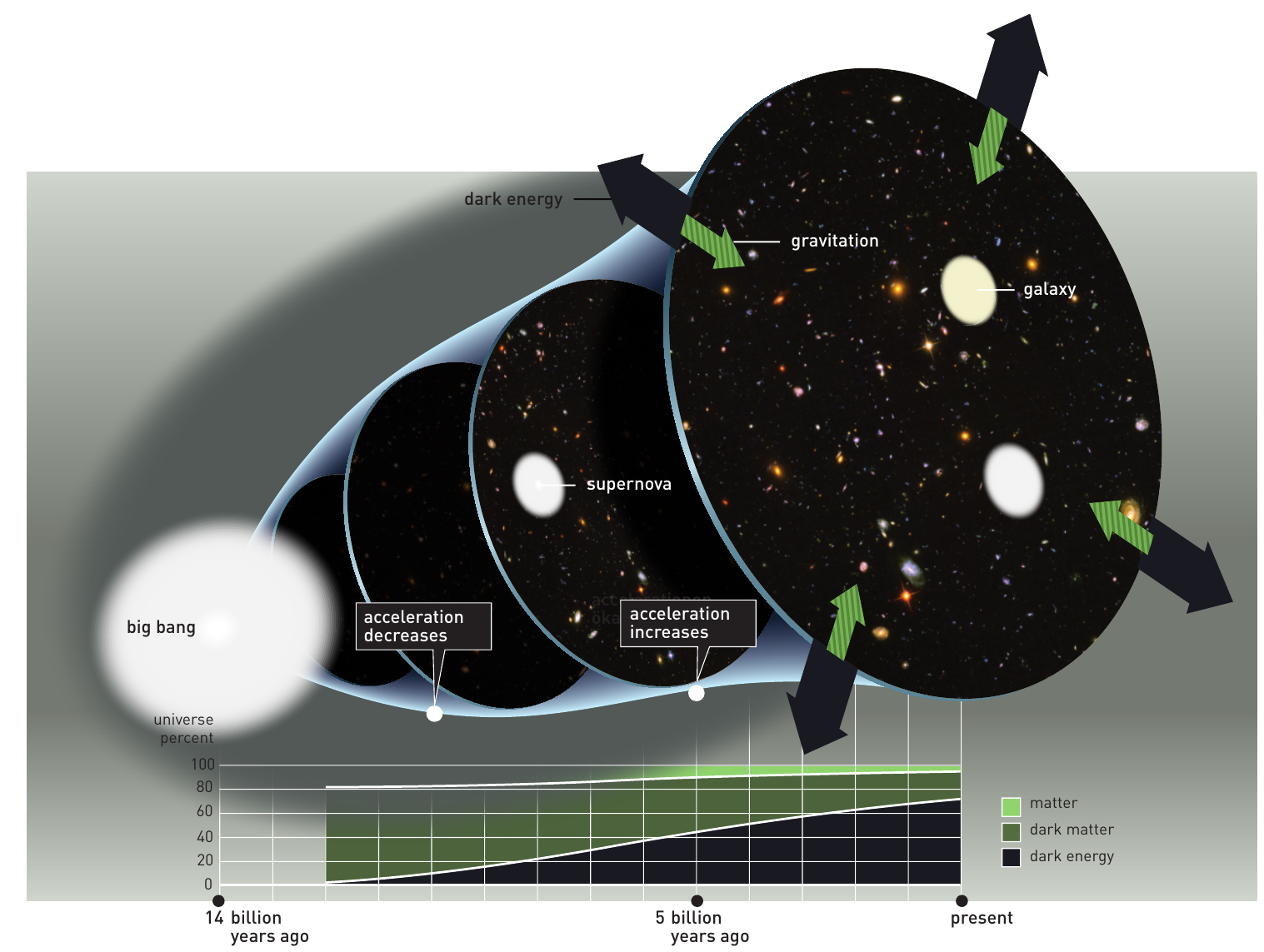}
\caption{
\small{
Expanding universe. Measurements of the Doppler shift in the light of supernova explosions \cite{Supernovae1,Supernovae2} were sufficiently precise to infer the second derivative of the expansion factor of the universe: the cosmic acceleration. The result was a surprise. Normally, matter is gravitationally attractive, causing a negative gravitational acceleration, but a uniform repulsive background turned out to dominate the net acceleration when the matter had become sufficiently diluted. This repulsive background is described by Einstein's cosmological constant $\Lambda$ and has been called ``dark energy'' for want of a more illuminating term. About $80\%$ of matter is equally elusive and has been called ``dark matter''. This paper makes a case for explaining dark energy as a consequence of vacuum fluctuations, showing how to overcome the problems \cite{Weinberg} that had plagued such an explanation since 1968.}
\label{expansion}}
\end{center}
\end{figure}

Einstein \cite{Einstein} introduced the cosmological constant $\Lambda$ in 1917 for having a static universe as solution of his equations of the gravitational field \cite{LL2}. What was the problem? The gravity of matter is attractive, so the matter distribution the universe is made of may first expand and then collapse --- like a stone thrown on Earth first rises and then falls, or it may expand forever, like a spacecraft on a voyage into space. In order to have an equilibrium, a repulsive force was required that acts on cosmological scales, but is otherwise too small to play a significant role. This repulsive background was provided for by Einstein's cosmological term \cite{Einstein}. Hubble's measurements \cite{Hubble} of the Doppler shift in the light coming from galaxies revealed a different picture however: the universe has been expanding at a positive rate. There was no need for an equilibrium solution. In the 1990s measurements on the light coming from certain supernovae \cite{Supernovae1,Supernovae2} became precise enough to calculate the second derivative of the expansion, which turned out to be positive, too (Fig.~\ref{expansion}). This implies that there is indeed a repulsive component to the gravitational force (more on this in Sec.~2.1). Recent measurements on the Cosmic Microwave Background (CMB) \cite{CMBPlanck} have established that the repulsive component takes the form of Einstein's cosmological term with constant $\Lambda$. The latest measurements on bright stars in galaxies \cite{Cepheids} seem to indicate, however, that $\Lambda$ is not constant, but has varied from the time the CMB was created to the present era. In any case, there is clear empirical evidence for the existence of the cosmological term and there are good quantitative measurements of $\Lambda$. 

The state of the theory could hardly be more different. The standard prediction of $\Lambda$ from quantum field theory deviates from the observed value by 120 orders of magnitude \cite{Weinberg}. This state of affairs is reflected in the contemporary scientific term for $\Lambda$: dark energy. The term conjures up a picture of some dark force filling the universe, while in reality it just says that current theoretical physics is unable to shed light on what ``dark energy'' really is. While there are many attempts \cite{Weinberg,DarkEnergy} to derive the cosmological term, none have been fully convincing in the following sense. Physical theories do not just give quantitative explanations of measured facts, but they connect the facts across several different areas of empirical investigation. The larger the range of explanation the more valueable is the theory and the more likely it is that the theory captures a portion of the truth. The attempts of deriving the cosmological constant from theory seem to fall into two categories: either are they off by many orders of magnitude, if they originate from other known physics, or they are highly specialized and therefore highly speculative. Yet a physically well-motivated explanation \cite{Zeldovich} has been put forward right from the beginning. 

Zel'dovich \cite{Zeldovich} suggested in 1968 that Einstein's cosmological term \cite{Einstein} comes from the physics of the quantum vacuum \cite{Milonni}. The quantum vacuum is the ground state of the physical fields --- the electromagnetic field and the fields of the weak and strong interaction. Fields are thought of being spanned by modes of harmonic oscillators, and the ground state of a harmonic oscillator is known to carry a zero--pointy energy and to appear with a fluctuating amplitude. An antenna put in vacuum would pick up noise that could be attributed to the fluctuating amplitudes of the modes it samples. The physics of the quantum vacuum has been well--tested and explains a huge set of phenomena, from the stickiness of materials on the nano scale to the limit trees can grow \cite{Koch}. Recent precision measurements of vacuum forces \cite{Rodriguez} including measurements of repulsive forces \cite{Levitation,CasimirEquilibrium} have tested the theory with an accuracy only limited by the knowledge of the material data involved. Zel'dovich's proposal \cite{Zeldovich} thus connects the cosmological constant to a range of completely different phenomena, which would give great strength to his theory. Unfortunately, while Zel'dovich derived the correct structure of the cosmological term, the predicted quantitative value for $\Lambda$ is in stark disagreement with the facts. 

The facts are that the universe is indeed dominated by vacuum fluctuations, simply because it is rather empty, and that the spectrum of these fluctuations is almost unlimited, because of the equivalence principle of general relativity \cite{LL2}. The emptiness of space is quantified in the number for the average mass density of space on cosmological scales at the present time:
\begin{equation}
\rho\approx 10^{-27} \mathrm{g}/\mathrm{cm}^{3} \,.
\label{density}
\end{equation}
The equivalence principle \cite{LL2} states that gravity acts equally on all bodies and on all scales. In particular, it implies that gravity acts equally on all fields on all wavelengths, down to the Planck scale where --- presumably --- classical general relativity \cite{LL2} does no longer hold. The equivalence principle has been well--tested, although not to the Planck scale, but it is a principle that connects a wide range of gravitational phenomena and therefore seems to capture a significant portion of the truth. The potential limitation of the principle is characterized by two numbers worth remembering, the Planck length $\ell_\mathrm{p}$ and the Planck mass $m_\mathrm{p}$ with the expressions
\begin{equation}
\ell_\mathrm{p} = \sqrt{\frac{\hbar G}{c^3}} \approx 1.6 \times 10^{-33} \mathrm{cm} \,,\quad
m_\mathrm{p} = \sqrt{\frac{\hbar c}{G}} \approx 2.1 \times 10^{-5} \mathrm{g} \,.
\label{planck}
\end{equation}
These are the length and the mass units one can construct by taking combinations of the fundamental constants of nature, Planck's $\hbar$, Newton's gravitational constant $G$ and the speed of light $c$. Their precise physical meanings are not known yet, but one may get some insight from the thermodynamics of causal horizons \cite{Jacobson} (generalizing black--hole thermodynamics \cite{Bekenstein}). Here the entropy increment is given by the area element divided by $4\ell_\mathrm{p}^2$, which suggests that $2\ell_\mathrm{p}$ plays the role of a fundamental length scale in gravitational physics. The Planck mass follows from the classical physics of black holes \cite{LL2} as the mass of a black hole of Schwarzschild radius $2\ell_\mathrm{p}$. It would be larger by a factor of $\sqrt[4]{5120 \pi} \approx 10$ than the minimal mass required for a quantum--mechanically stable black hole \cite{LeoBook} (that is not immediately evaporated \cite{Hawking}). Vacuum modes with wavelengths comparable to the Planck length, if they were to exist, would thus immediately generate black holes and could no longer be taken as passive objects obeying the equivalence principle, but as active agents dissolving the structure of space--time. 

Let me give a simple argument, based on dimensional analysis, why the numbers (\ref{density}) and (\ref{planck}) appear to be in conflict with each other. The energy density $\varepsilon_\mathrm{vac}$ of the quantum vacuum must be proportional to $\hbar$, because it is made by the zero--point energy of all the field modes combined that goes with $\hbar$. In order to get an energy one should multiply $\hbar$ by $c$ and divide by the forth power of a length. This length would correspond to the minimal wavelength of the field modes, which one may take as the Planck length. This gives
\begin{equation}
\varepsilon_\mathrm{vac} \sim \frac{\hbar c}{\ell_\mathrm{p}^4} \,.
\label{prediction}
\end{equation}
Converted into a mass density $\rho_\mathrm{vac} = \varepsilon_\mathrm{vac}\,c^{-2}$ one gets
\begin{equation}
\rho_\mathrm{vac} \sim \frac{m_\mathrm{p}}{\ell_\mathrm{p}^3} 
\end{equation}
and hence, using the numerical values (\ref{density}) and (\ref{planck}) for the actual mass density $\rho$ and the Planck length and mass: 
\begin{equation}
\log_{10} (\rho_\mathrm{vac}/\rho) \approx 120 \,.
\end{equation}
The elementary prediction (\ref{prediction}) of quantum field theory disagrees with astronomical observations on a truly astronomical scale.  Resolving this conflict between theory and reality, while keeping Zel'dovich's connection between the cosmological constant and the quantum vacuum, requires a re--examination of the empirical evidence and the supporting theory, as follows. 

\section{The evidence}

\subsection{Cosmology}

Astronomical observations have shown \cite{Survey} that the universe is homogeneous and isotropic on cosmological scales (larger than $100 \mathrm{Mpc}$). This implies \cite{LL2} that infinitesimal spatial distances can only change with a universal factor $n$ that is uniform in space but may depend on time $t$. In the spirit of optical analogues of gravity \cite{Gordon,Plebanski,Schleich,LeoPhil} we may regard the factor $n$ as a uniform refractive index that varies in time. The matter and energy in the universe at large acts, on average, as a fluid \cite{LL6}. This fluid must be at rest relative to the expanding universe, for otherwise the direction of its movement would violate the condition of isotropy. The fluid is therefore solely characterized by its energy density $\varepsilon$ and pressure $p$ where the energy density is related to the mass density by $\varepsilon=\rho c^2$. These postulates, supported by empirical evidence \cite{Survey}, constitute the cosmological principle \cite{LL2}.

One may formulate the cosmological principle in terms of a space--time metric \cite{LL2} and deduce from Einstein's equations the Friedmann equations of cosmic evolution \cite{LL2}. It is however possible --- and instructive --- to deduce the laws of cosmology almost exclusively from Newtonian physics \cite{Milne}. Take one point in space and imagine a sphere around this point filled with the cosmic fluid of uniform mass density $\rho$. Consider the Newtonian gravitational potential inside the sphere. Gauss' theorem implies that it is the potential of a harmonic oscillator with spring constant given by Newton's constant $G$ times the mass density $\rho$ multiplied by the volume $4\pi/3$ of the unit sphere. Relativity \cite{LL2} makes only one correction: in addition to the mass density the pressure also causes gravity. To be precise, relativity adds the term $3p/c^2$ to $\rho$ in the effective mass density of gravity in cosmology. One gets from the equation of motion of the harmonic oscillator:
\begin{equation}
\frac{\ddot{n}}{n} = - \frac{4\pi G}{3} \left(\rho+\frac{3p}{c^2}\right) ,
\label{newton}
\end{equation}
which is known as Newton's equation of the universe \cite{Telephone}. In the case when the pressure is much smaller than the rest energy density the gravity of matter thus establishes a restoring force with negative acceleration $\ddot{n}$. The observation \cite{Supernovae1,Supernovae2} of $\ddot{n}>0$ in the present era reveals however that there is a substantial negative pressure with $p<-\rho c^2/3$. The analysis of CMB fluctuations \cite{CMBPlanck} identifies a constant contribution of $\varepsilon_\Lambda$ to the total energy density $\varepsilon$ with
\begin{equation}
p_\Lambda = -\varepsilon_\Lambda \,.
\label{lambda}
\end{equation}
This constant negative pressure combined with the constant positive energy density $\varepsilon_\Lambda$ acts like a uniform source of repulsive gravity --- as if the matter of the universe were embedded in a uniform background of opposite gravitational charge \cite{Kolomeisky}. With increasing expansion the mass density of matter decreases such that this uniform background becomes increasingly important. At the present time $\varepsilon_\Lambda/c^2$ amounts to about $70\%$ of the total mass density (\ref{density}). This constant background is called dark energy. More prosaically --- and more accurately --- it corresponds to the cosmological constant introduced by Einstein \cite{Einstein}. 

One more ingredient is needed for completing the laws of cosmic evolution, and that comes from thermodynamics \cite{LL5}. The cosmic fluid is assumed to be isentropic --- entropy is conserved, such that a change in the total energy $E$ of a volume $V$ is given by $\mathrm{d}E=-p\,\mathrm{d}V$. As the volume $V$ varies with $n^3$ and $E=\varepsilon V$ one gets the second Friedmann equation \cite{LL2}:
\begin{equation}
\dot{\varepsilon} = -3(\varepsilon + p) H
\label{f2}
\end{equation}
where $H$ denotes the Hubble constant
\begin{equation}
H=\frac{\dot{n}}{n} \,.
\label{hubble}
\end{equation}
Note that $H$ is not actually constant in general, except when the universe is exponentially expanding (in the case of de Sitter space \cite{deSitter}). 

Having established the equations of motion for the expanding universe, Eqs.~(\ref{newton}) and (\ref{f2}), we integrate them. Consider the quantity $K$ defined by the relation 
\begin{equation}
H^2 + \frac{K}{n^2} = \frac{8\pi G}{3c^2}\, \varepsilon
\label{f1}
\end{equation}
with $\varepsilon=\rho c^2$. Differentiating Eq.~(\ref{f1}) and making use of the Newton equation (\ref{newton}) for $\dot{H}+H^2=\ddot{n}/n$ as well as the thermodynamic relation (\ref{f2}) and the undifferentiated Eq.~(\ref{f1}) reveals that $\dot{K}=0$. In other words, $K$ is a constant. It turns out in general relativity \cite{LL2} that $K$ describes the spatial curvature; $K$ is positive for positively curved space and negative for negative spatial curvature, and in the marginal case of flat space
\begin{equation}
K = 0 \,.
\label{flat}
\end{equation}
Newtonian cosmology does account for the curvature of space in principle \cite{MM}. The analysis of CMB data \cite{CurveMeas} has shown that in practice space is approximately flat on cosmological scales such that Eq.~(\ref{flat}) holds to a good approximation. Equation (\ref{f1}) is known as the first Friedmann equation \cite{LL2}, which concludes all the cosmology we will need in this article. 

\subsection{Vacuum}

\begin{figure}[h]
\begin{center}
\includegraphics[width=20pc]{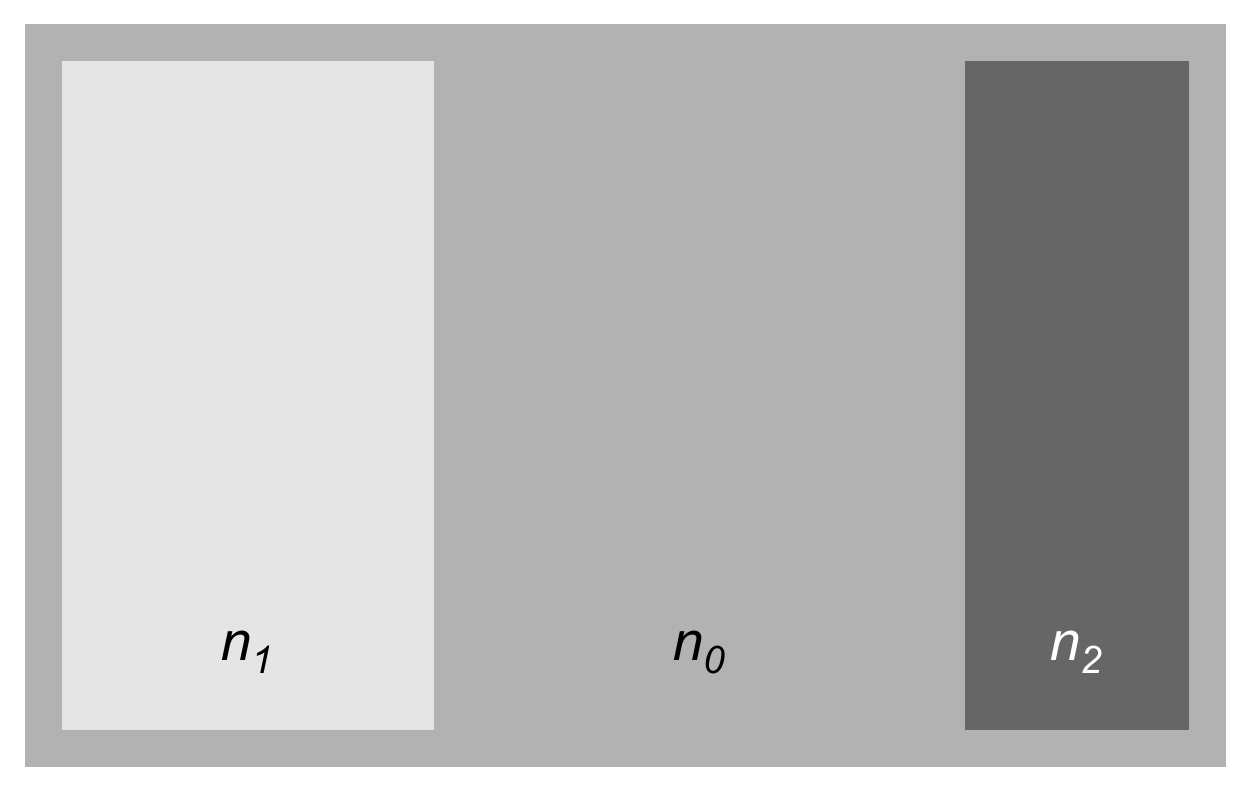}
\caption{
\small{
Typical Casimir setup. Two dielectric bodies with refractive indices $n_1$ and $n_2$ are immersed in a dielectric background of refractive index $n_0$. The Casimir effect of the quantum vacuum creates a mechanical force density at the interfaces of the bodies. Elsewhere the force is zero. Renormalization is needed in order to calculate the force. In the simplest procedure, the bare vacuum energy is renormalized by the energy one obtains when moving the bodies within the background to infinity. Taking the difference between the two gives a finite result that agrees with experiments. 
}
\label{typical}}
\end{center}
\end{figure}

Having collected the evidence from cosmology, consider now the case for forces of the quantum vacuum \cite{Forces}. Without exception, the empirical evidence for vacuum forces comes from Atomic, Molecular and Optical Physics (AMO) and here from quantum fluctuations of the electromagnetic field. They appear as the van der Waals and Casimir forces \cite{Rodriguez}. The Casimir force is typically (Fig.~\ref{typical}) a force between electrically neutral bodies of refractive indices $n_i$ immersed in a uniform background with index $n_0$. Strictly speaking \cite{LL8}, dielectrics are characterized by the refractive index $n$ and the impedance $Z$; here we assume $Z=1$ for simplicity. At sufficiently low temperatures (for thermal wavelengths larger than the characteristic distances) the Casimir force originates from fluctuations of the quantum vacuum \cite{LL9}. These are in general not fluctuations of the electromagnetic field in empty space, but inside the media the bodies and the background are made of. The term ``vacuum'' is used to state that they are quantum fluctuations of the field in the ground state, given the arrangement of dielectrics. These fluctuations carry energy and exert stress that does mechanical work; the divergence $\nabla\cdot\sigma$ of the stress tensor $\sigma$ gives the force density. For an arrangement of dielectric bodies of uniform refractive indices in a uniform background (Fig.~\ref{typical}) the force density is entirely concentrated at the surfaces of the bodies, causing the Casimir force \cite{Casimir}.

There are various theories for the Casimir effect of the quantum vacuum. Casimir's original theory \cite{Casimir} considers the total sum $E$ of all the ground state energies $\hbar\omega_m/2$ of all electromagnetic modes involved. This theory presumes the existence of stationary modes oscillating with the circular frequencies $\omega_m$. In realistic media the concept of modes become questionable, because real materials are dissipative such that stationary modes no longer exist, strictly speaking. The theory explaining Casimir forces in realistic dielectrics, Lifshitz theory \cite{LL9,Lifshitz,LDP,Scheel} takes a different starting point: the fluctuation--dissipation theorem \cite{Scheel}. The theorem allows to express the energy density and stress tensor \cite{Pita,Burger} in terms of the electromagnetic Green functions. Lifshitz theory \cite{LL9,Lifshitz,LDP,Scheel} has become the theoretical workhorse in AMO Casimir physics \cite{Rodriguez}. There it agrees with precision experiments \cite{Levitation,CasimirEquilibrium} with an accuracy in the percent level that appears to be only limited by the spectral range and accuracy of the material data needed for calculating the force. 

The theory of Casimir forces, whether it is Lifshitz theory \cite{LL9,Lifshitz,LDP,Scheel} or any other formulation \cite{Milonni,BKMM} involves one crucial step: renormalization. Without it the energy density and the stress of the quantum fluctuations are infinite. To see this in its most elementary form, consider a one--dimensional toy model, a waveguide of length $a$ with reflecting end cups. The spectrum this one--dimensional cavity supports is given by the discrete circular frequencies $\omega_m = (\pi c/a) m$ with positive integers $m$. One thus gets for the density of the total zero--point energy 
\begin{equation}
\varepsilon_{\mathrm{\small 1D}} = \frac{E}{a} = \frac{1}{a}\sum_{m=1}^\infty \frac{\hbar\omega_m}{2} = \frac{\hbar c}{a^2} \frac{\pi}{4}\sum_{m=1}^\infty m = \infty \,.
\label{toy}
\end{equation}
Renormalization extracts the part of the energy and stress that does mechanical work. In the case of the one--dimensional toy model of Eq.~(\ref{toy}) one obtains \cite{BKMM} the renormalized vacuum energy density
\begin{equation}
\varepsilon_{\mathrm{vac  1D}} =  -\frac{\pi}{48} \,\frac{\hbar c}{a^2}  
\end{equation}
as if the sum of all positive integers were effectively $-1/12$. In its most physically convincing and numerically efficient form \cite{Reid} the renormalization is done by comparing the energy of the arrangement of dielectric bodies at finite distances with the energy they had if they were infinitely far apart. It is important to keep the immersion at the same index $n_0$, for otherwise the difference in energies would diverge. This simple renormalization procedure is consistent with Lifshitz theory \cite{Reid} and hence has been well--tested by precision experiments. 

In cosmology, the expansion factor $n$ plays the role of the refractive index. Here $n$ is uniform in space but varies in time, whereas in the arrangement of dielectric bodies $n$ varies between the bodies and the background --- it varies in space and not in time. The energies and stresses of vacuum fluctuations should be just the same though, as they are both given by the energy--momentum tensor of the electromagnetic field (in the presence of the gravitational field or in media). What differs is only the way energies and stresses act. In AMO physics, the divergence of the stress produces directly the force density. In cosmology, the energy density and pressure contribute to the cosmic expansion due to their gravity, as described in the Friedmann equations (\ref{f2}) and (\ref{f1}). Yet in general relativity the entire energy and stress is supposed to gravitate, and not just the renormalized part, which produces a figure that disagrees with fact by 120 orders of magnitude, as shown in Sec.~I. In view of this obvious conflict with reality, why should we not take the empirical fact of the astronomically small mass density, Eq.~(\ref{density}), as evidence that the bare vacuum energy, for whatever reason, does not gravitate? 

\section{The theory}

\begin{figure}[h]
\begin{center}
\includegraphics[width=20pc]{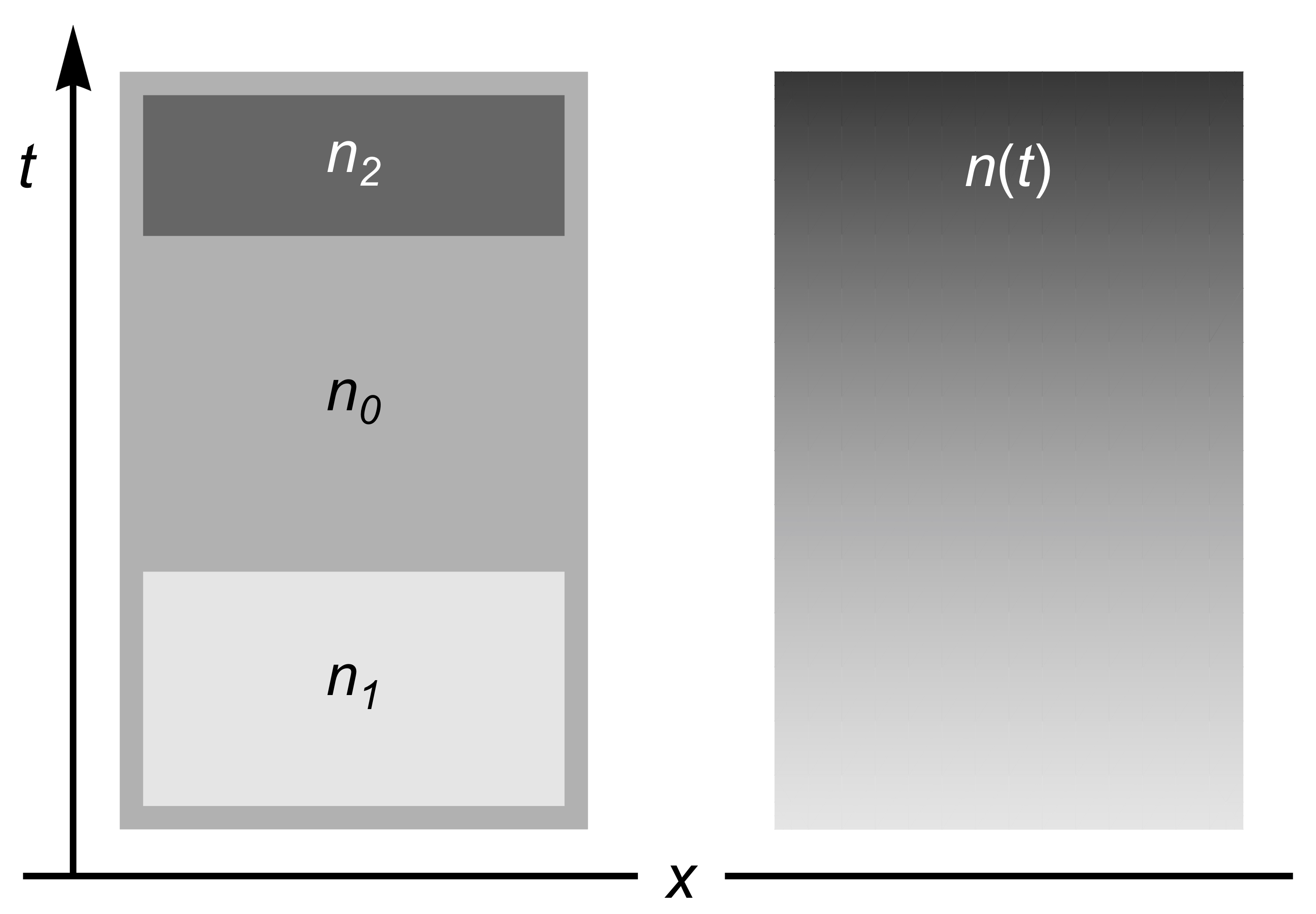}
\caption{
\small{
Casimir effect in time instead of space (Fig.~\ref{typical}). Space--time diagrams of dielectrics with time--dependent refractive indices indicated by different shades of grey. Left: discrete changes of the refractive index similar to the typical Casimir setup (Fig.~\ref{typical}). Right: continuously changing index profile $n(t)$. In the latter case the Casimir effect creates a continuous force density. 
}
\label{rot}}
\end{center}
\end{figure}

The bare vacuum energy and stress has not appeared in any experimental test of Casimir forces, nor does it appear in cosmology. Should we still take it as real? Or should we rather regard it as an artefact of the theory? Assume that renormalization is real, that the bare vacuum and stress always needs to be removed, even in gravity. The universe would still create a Casimir energy $\varepsilon_\mathrm{vac}$ and pressure $p_\mathrm{vac}$, because it is evolving. Instead of changing in space (on cosmological scales) the expansion factor $n$ changes in time, as if the spatial refractive index profile (Fig.~\ref{typical}) would be turned into a space--time diagram (Fig.~\ref{rot}). This time--dependent refractive--index profile produces a renormalized $\varepsilon_\mathrm{vac}$ and pressure $p_\mathrm{vac}$ that will be significantly smaller than the notorious 120 orders of magnitudes of the bare theory \cite{Weinberg}.

Yet there is one important subtlety to consider: the cosmic expansion $n(t)$ is continuous, it does not happen in jumps from one constant $n$ to the next, whereas in the typical setting for the Casimir effect (Fig.~\ref{typical}) the refractive index $n$ is piecewise homogenous. The Casimir force in inhomogeneous media is no longer concentrated at the surfaces of dielectric bodies --- because there are no distinct bodies one could distinctly identify in gradually varying media. The force density will be a continuous function. One might think \cite{Simpson} one could simply discretise a continuous variation of $n$ and then calculate the force at each discrete transition from one $n$ to the next. This discretisation procedure would create force densities localized at the interfaces that, in the continuum limit, approach a smooth force distribution. It has been one of the surprises in theoretical Casimir physics \cite{SimpsonSurprise} that this simple discretisation of the refractive index does not work \cite{Simpson}. It then took some time and effort to find a workable solution \cite{Grin1} for the case of planar media where $n$ varies in one direction in space. 

The problem is that the physically intuitive renormalization procedure of the standard Casimir effect (Fig.~\ref{typical}) is not applicable anymore. In a continuously varying medium it is simply impossible to move the bodies of the medium to infinity, for working out their bare Casimir energy, because the medium consists of {\it one} indivisible body. A local renormalization procedure is needed. For finding this procedure, insight and intuition comes from Schwinger's source theory \cite{SchwingerSource}. Schwinger assumed here \cite{SchwingerSource} that the quantum fluctuations of fields originate from quantum fluctuations of their sources. In the case of dielectric media, these sources are the electric and magnetic polarizations of the dielectrics. Their response to the electromagnetic field forms the refractive--index profile. According to the dissipation--fluctuation theorem \cite{Scheel} the polarizations fluctuate, even at zero temperature when they are purely quantum. The fluctuating charges and currents of the medium create electromagnetic fields that propagate them in time across space. The fields are the messengers, not the sources of the quantum fluctuations. This picture agrees with the modern quantization procedure \cite{KSW,PhilbinQ1,PhilbinQ2,Buhmann,Horsley,HorsleyPhilbin} of the electromagnetic field in dispersive and dissipative media. It also agrees with the starting point of Lifshitz theory \cite{Lifshitz}: Rytov's fluctuating sources \cite{Rytov}. The picture suggests a local renormalization procedure as follows.

Each point of the medium is mentally split into two points, one is the emitter and the other the receiver of electromagnetic fields. Driven by quantum fluctuations, the emitter sends out an electromagnetic wave. The wave propagates, is partially reflected in the medium and is then picked up by the receiving end of the cell. Of course, all the other cells of the medium are also emitting electromagnetic waves the receiver is responding to. But averaged over the fluctuations, each cell can only interact with the field emitted by itself, because the different cells of the medium are not correlated, their emissions are independent from each other and random. Only the wave emitted by one and the same cell has a well--defined phase relation with itself and so, on average, only this field will have a physical effect onto the cell. The emitted field consists of two part, an outgoing wave and a scattered wave that is reflected at the inhomogeneities of the medium. The receiver does only interact with the scattered wave. Yet the theoretical description of the field does also contain the outgoing wave. In order to describe the idea that the receiver interacts only with the scattered radiation, one should subtract in the energy and stress the contribution coming from the outgoing wave. This subtraction procedure of the unphysical interaction of the point with itself corresponds to the renormalization procedure in Schwinger's picture \cite{Grin1}. It is implemented in Lifshitz theory \cite{LL9,Lifshitz,LDP,Scheel} by subtracting the outgoing Green function from the total Green function. This is how renormalization was conceived in Lifshitz', Pitaevskii's and Dzyaloshinskii's original theory \cite{LDP}, whereas the intuitive idea of comparing the Casimir energy between dielectric bodies at finite and infinite distances was Casimir's \cite{Casimir}. Casimir's method has made it into Lifshitz theory in its modern numerical implementation \cite{Reid}, whereas the original renormalization procedure in Lifshitz theory has been the tool for analytical calculations. 

Now, the outgoing wave emitted at a given point in the medium depends on the dielectric environment of that point: it depends on the local refractive index $n$. Without taking the local $n$ into account, the difference between the total and the bare stress would not converge to a finite value. Lifshitz theory with the original renormalization procedure \cite{LDP} agrees with experiments \cite{Levitation,CasimirEquilibrium} where the dielectric backgrounds differ from vacuum ($n_0\neq 1$). This gives experimental evidence supporting the concept of local renormalization and ruling out any idea that renormalization amounts to subtracting the large, global energy of the bare quantum vacuum. Renormalization is local, but how local is it? The outgoing wave needs at least one cycle of oscillation for establishing itself as a wave --- the oscillating magnetic field made by the source current induces an electric field that, in turn, drives the magnetic field, which induces an electric field and so on. Therefore it seems plausible that the outgoing wave depends not only on the local value of $n$, but also on the first two derivatives of $n$. For piece--wise homogeneous media (Fig.~\ref{typical}) the dependence of the renormalization on derivatives of $n$ does not matter, as those derivatives are zero. For inhomogeneous media, the appearance of the extra derivatives explains why the naive discretisation of the medium fails to give a converging Casimir force \cite{Simpson}. For planar media, the renormalization to second order was proven \cite{Grin1} to converge. The theory \cite{Grin1} has not been tested in experiments yet, but it predicts effects \cite{Grin2} that seem measurable with current experimental techniques, and hence are testable.

The theory \cite{Grin1} of the Casimir stress inside inhomogeneous planar materials makes one more prediction that, when widely extrapolated to cosmological scales, explains why the Casimir effect might play a role in cosmology: the convergence of the renormalization relies on dispersion. Ordinary dielectric materials are dispersive in the sense that the refractive index $n$ depends on frequency. The Casimir effect is a broadband electromagnetic phenomenon \cite{Rodriguez} depending on the entire frequency window of the material. For large frequencies all materials become completely transparent, $n\rightarrow 1$. Without this feature, the renormalized stress would contain a logarithmically diverging contribution \cite{Grin1}. On the other hand, the ``material'' of general relativity  --- the geometry of space and time --- acts on all frequencies equally, as a consequence of the equivalence principle \cite{LL2}. Therefore even the renormalized $\varepsilon_\mathrm{vac}$ would still diverge, although significantly weaker than the unrenormalized one. The wavelength range contributing to the forces of the quantum vacuum would go to the Planck length (\ref{planck}) where, presumably, the equivalence principle ceases to hold. So $\varepsilon_\mathrm{vac}$ would not grow with the inverse forth power of the Planck length as in Eq.~(\ref{prediction}) that produces the wrong 120 orders of magnitudes, but significantly weaker. The logarithmic divergence \cite{Grin1} is not sufficient though, for the following reason.

The divergence of the vacuum energy density and stress with the cutoff scale occurs only in inhomogeneous media. The prefactor of the divergence is therefore not a universal constant, but vanishes for constant $n$. The prefactor must depend on derivatives of $n$. In cosmology, $n$ varies on the time scale of the inverse of the Hubble constant $H$. For being able to influence the cosmic evolution described by the Friedmann equation (\ref{f1}), the energy density $\varepsilon_\mathrm{vac}$ should go like $H^2$. Being a quantum energy $\varepsilon_\mathrm{vac}$ must be proportional to $\hbar$; having the physical dimensions of an energy density suggests that  $\varepsilon_\mathrm{vac}$ should go like $(\hbar/c) H^2/\ell^2$. Indeed, identifying $\ell$ with the Planck length (\ref{planck}) satisfies the Friedmann equation (\ref{f1}) for $K=0$. This simple dimensional analysis \cite{Tractatus} indicates that $\varepsilon_\mathrm{vac}$ must diverge with an inverse square length for having a case for the Casimir effect in cosmology.

In planar media \cite{Grin1} the refractive--index profile depends only on one direction of space (Fig.~\ref{typical}), in spatially--flat cosmology it depends only on time (Fig.~\ref{rot}). This apparently simple rotation of the spatial profile into a space--time diagram involves two important subtleties that distinguishes the cosmological case from the planar case: horizons and causality. Horizons are a consequence of Hubble's law \cite{LL2}: the space around a point appears to expand with a velocity that grows with the Hubble constant (\ref{hubble}) times the distance. Hubble's law is a simple kinematic feature of spatially uniform expansion where distances $r$ grow as $n(t)r$. Differentiation shows that a distant point moves away with velocity $H$ times the distance $nr$. This apparent velocity may become arbitrarily large. At the horizon the expansion speed is equal to the speed of light such that the interior of the horizon is causally disconnected from the rest of the universe. No wave from outside of the horizon can reach the point. The cosmological horizon is completely analogous \cite{Tractatus} to the event horizon of the black hole \cite{Brout}. In particular, the horizon is predicted \cite{Tractatus,GibbonsHawking} to emit the analogue of Hawking radiation \cite{Hawking} with temperature
\begin{equation}
k_\mathrm{B} T = \frac{\hbar H}{2\pi}
\label{gh}
\end{equation}
where $k_\mathrm{B}$ denotes Boltzmann's constant. Although the temperature (\ref{gh}) lies significantly below the $2.7\mathrm{K}$ of the CMB ($2\times 10^{-29}\mathrm{K}$ for the current inverse Hubble constant of about $10^{10}$ years) it turns out \cite{Tractatus} to play a dominant role in the regularization of the vacuum energy. 

Another extremely subtle but crucial feature of the cosmological case is causality. Recall that in the local renormalization method each point is mentally split into two: the emitter and the receiver. While emitter and receiver can change places in space, they cannot alter their order in time: emission must precede reception. This subtle constraint from causality, combined with the radiation of the cosmological horizon, creates \cite{Tractatus} the cosmologically relevant divergence of the vacuum energy in the second--order renormalization procedure established for planar media \cite{Grin1}. The Casimir energy may become cosmologically relevant. But does it explain dark energy?

\section{The anomaly}

The characteristic feature, Eq.~(\ref{lambda}), of the cosmological constant is the negative pressure $p_\Lambda$ equal in magnitude to the positive energy density $\varepsilon_\Lambda$. Is this $\varepsilon_\Lambda$ the renormalized energy density $\varepsilon_\mathrm{vac}$ of the quantum vacuum? No, because the corresponding pressure $p_\mathrm{vac}$ must be $\varepsilon_\mathrm{vac}/3$. This is a consequence of the fact that the energy of an electromagnetic wave is equal to its momentum times the speed of light. The energy of a volume element filled with incoherent radiation must therefore be equal to the momentum transfer, {\it i.e.} the pressure, in the three directions of space, which gives $\varepsilon_\mathrm{vac}=3p_\mathrm{vac}$.

So where does $\varepsilon_\Lambda$ and $p_\Lambda$ come from? Consider the conservation of energy and momentum as expressed in the second Friedmann equation (\ref{f2}). With $p_\mathrm{vac}=\varepsilon_\mathrm{vac}/3$ one gets the differential equation $\dot{\varepsilon}_\mathrm{vac}=-4\varepsilon_\mathrm{vac}$ with the solution $\varepsilon_\mathrm{vac}\propto n^{-4}$. This would imply that $\varepsilon_\mathrm{vac}$ is only a function of the expansion factor $n$. Yet if $\varepsilon_\mathrm{vac}$ is a Casimir energy it must also depend on derivatives of $n$. The vacuum energy $\varepsilon_\mathrm{vac}$ violates energy--momentum conservation. Wald \cite{Wald} understood the root of the problem: the lack of reciprocity in the renormalization procedure. Recall the point--splitting picture: the emitter sends out fluctuating electromagnetic waves, the emitter receives the reflected waves. The outgoing waves depend on the refractive--index profile around the emitter, which differs from the profile around the receiver when $n$ varies. Emitter and receiver and not reciprocal. Wald realized that this lack of reciprocity violates the conservation of energy and momentum \cite{Wald}. Intuitively one may see this as a recoil imbalance \cite{Tractatus}: the recoil of the wave on the emitter is not the same as the recoil on the receiver.

Consider the energy density of the recoil imbalance --- just add a term $\varepsilon_\mathrm{recoil}$ to $\varepsilon_\mathrm{vac}$. The term should describe exactly the missing energy in the energy--momentum balance, {\it i.e.}\ on the left--hand side of the Friedmann equation (\ref{f2}). This is only possible if the right--hand side is not changed by the associated pressure. The recoil pressure must therefore be the exact opposite of the recoil energy, precisely as for $\varepsilon_\Lambda$ and $p_\Lambda$ in Eq.~(\ref{lambda}). One may thus identify $\varepsilon_\Lambda$ with $\varepsilon_\mathrm{recoil}$ and arrives at a physical picture for the energy of the cosmological constant --- dark energy. The energy and pressure of $\Lambda$ is created in the recoil imbalance of vacuum fluctuations emitted and received in the medium of space--time \cite{Tractatus}. The technical term for this imbalance is called trace anomaly \cite{Wald}. 

Not only gives this idea a physical picture for the cosmological constant $\Lambda$, it also establishes a method for calculating $\varepsilon_\Lambda$ (and $p_\Lambda=-\varepsilon_\Lambda$). Consider for completeness the rest of matter and radiation described by some energy density $\varepsilon_\mathrm{m}$ and pressure $p_\mathrm{m}$ in addition to the energy and pressure of the quantum vacuum. The total energy density and pressure is then given by
\begin{equation}
\varepsilon = \varepsilon_\mathrm{m} + \varepsilon_\mathrm{vac} + \varepsilon_\Lambda \,,\quad
p = p_\mathrm{m} + \frac{1}{3}\, \varepsilon_\mathrm{vac} - \varepsilon_\Lambda \,.
\end{equation}
Assume for simplicity and in agreement with observation \cite{CurveMeas} that the universe is spatially flat. Differentiating the first Friedmann equation (\ref{f1}) and using the second Friedmann equation (\ref{f2}) establishes an evolution equation of the universe: 
\begin{equation}
\dot{H} = -\frac{8\pi G}{c^2} \left(\varepsilon_\mathrm{m} + p_\mathrm{m} + \frac{4}{3}\,\varepsilon_\mathrm{vac}\right)
\label{evolution}
\end{equation}
similar to the Newton equation (\ref{newton}). Given $\varepsilon_\mathrm{m}$ and $p_\mathrm{m}$ as functions of $n$ and $\varepsilon_\mathrm{vac}$ as functions of $n$ and its derivatives, Eq.~(\ref{evolution}) defines an equation of motion for the universe on cosmological scales. This equation of motion is independent of the cosmological term $\varepsilon_\Lambda$. Given a solution of the cosmic dynamics, the energy density $\varepsilon_\Lambda$ follows from the Friedmann equation (\ref{f1}) with $\varepsilon = \varepsilon_\mathrm{m} + \varepsilon_\mathrm{vac} + \varepsilon_\Lambda$. 

It remains to calculate $\varepsilon_\mathrm{vac}$ for spatially uniform refractive index profiles $n$ changing in time $t$. Assuming exactly the same renormalization  as for refractive index profiles changing in one direction of space \cite{Grin1} but taking into account the temperature (\ref{gh}) of cosmic horizons and causality in the point--splitting method produces after renormalization the vacuum energy density \cite{Tractatus} 
\begin{equation}
\varepsilon_\mathrm{vac} = - \frac{\hbar}{12\pi^2c}\,\frac{\Delta}{\ell^2}
\label{result}
\end{equation}
where $\Delta$ depends on the time derivatives of the Hubble constant $H$ and has units of $H^2$. Assuming in the renormalization procedure that the horizon temperature should be taken with respect to conformal time \cite{Tractatus}, gives \cite{Tractatus} 
\begin{equation}
\Delta = \partial_t^3\frac{1}{H} + H \partial_t^2\frac{1}{H} \,.
\label{resultdelta}
\end{equation}
The energy density (\ref{result}) goes with the inverse square of the cutoff length $\ell$. Setting this length to the order of the Planck length (\ref{planck}) gives an $\varepsilon_\mathrm{vac}$ that contributes at the right level to the cosmic evolution described by Eq.~(\ref{evolution}). This vacuum energy is neither too large nor too small, neither is it off by the notorious 120 orders of magnitude nor is it as insignificant as one might expect for the Casimir force. Note that Eqs.~(\ref{result}) and (\ref{resultdelta}) hold not only in a spatially flat cosmology, but also in homogeneous and isotropic spaces with curvature \cite{Tractatus}. The calculation \cite{Tractatus} was done for electromagnetic fields. Assuming the same principal behaviour for the other fields of the Standard Model would amount to multiplying Eq.~(\ref{result}) with the number of the independent field components divided by two (the polarizations of the electromagnetic field). As the precise cutoff length is not precisely known, one can express both the multitude of fields and the cutoff in an effective constant prefactor $\eta$ and write Eq.~(\ref{result}) as
\begin{equation}
\varepsilon_\mathrm{vac} = - \frac{\hbar\eta}{12\pi^2c}\,\frac{\Delta}{\ell_\mathrm{p}^2} \,.
\label{eta}
\end{equation}
The following picture emerges from the theory \cite{Tractatus}. The cosmic evolution, described by the scaling factor $n$ and its derivatives, generates the energy density (\ref{eta}) of the quantum vacuum. The vacuum energy acts back on the cosmic dynamics as described in the evolution equation (\ref{evolution}). Consistent with this evolution is the effective cosmological constant with energy density $\varepsilon_\Lambda$ given by the Friedmann equation (\ref{f1}) with $\varepsilon = \varepsilon_\mathrm{m} + \varepsilon_\mathrm{vac} + \varepsilon_\Lambda$. The cosmological constant corresponds to a trace anomaly \cite{Wald} one may interpret as a recoil imbalance of vacuum fluctuation in the ``material'' of space--time \cite{Tractatus}. The vacuum energy of Eqs.~(\ref{resultdelta}) and (\ref{eta}) vanishes if the Hubble constant (\ref{hubble}) is indeed constant, {\it i.e.} for exponential expansion. In this case, the cosmological constant $\Lambda$ is constant as well. Otherwise it will vary, which might explain the observed variation of $\Lambda$ \cite{Cepheids}. This picture shares some features with the theory of quintessence \cite{Caldwell} but it does not require any new fields, just new concepts in fields as old as quantum electromagnetism. 

\section{Conclusions}

There is no empirical evidence for the bare vacuum energy of fields. Neither does the bare vacuum energy do mechanical work \cite{Forces}, nor does it gravitate. Therefore it seems wise to take for the vacuum energy in the universe a renormalized energy density. Assuming the same renormalization procedure as for the Lifshitz theory in planar media \cite{Grin1} while taking into account cosmological horizons and causality, gives a vacuum energy density of the right order of magnitude \cite{Tractatus}. The resulting theory makes concrete predictions about the cosmic evolution, Eq.~(\ref{evolution}), apart from one parameter in Eq.~(\ref{eta}) that cannot be determined yet. The result seems encouraging, but of course it remains to be seen whether the theory reproduces the astronomical facts in detail, and not only their order of magnitude. The theory was developed \cite{Tractatus} for the electromagnetic field as carrier of vacuum fluctuations; it therefore remains to be checked whether it can be extended to other fields. 

Analogues of gravity \cite{Gordon,Plebanski,Schleich,LeoPhil} have played a decisive role in developing the theory \cite{Tractatus} and analogues may be important for testing crucial components of the theory in experiments. The theory extends the renormalization procedure of the Casimir stress in inhomogeneous planar media \cite{Grin1} to spatially uniform, time--dependent materials. Direct measurements of the Casimir force inside planar media are difficult, the easiest is perhaps a test of the extreme behaviour predicted in Ref.~\cite{Grin2}. But one could emulate Casimir forces with other fluctuations, as long as they have a similar spectrum than the vacuum fluctuations, replacing the $\hbar$ in the Casimir force by an effective noise parameter of significantly larger magnitude, which would enhance the effect. Measuring Casimir forces in time--dependent media is not simple either. Here also analogues may play an important role for crucial tests. 

While the physics of the Casimir effect of separate bodies is well--understood, the Casimir force inside materials has remained a fairly underdeveloped subject. There is enormous scope for research in both theory and experiment. If the Lifshitz theory of the cosmological constant \cite{Tractatus} does indeed agree with the astronomical facts in detail, either directly or after minor modifications, it would not only shed light on a rather dark subject in cosmology, but also motivate a better understanding of the forces acting on the nanoscale in the everyday world.

\newpage

\noindent
{\bf Funding.} European Research Council, Israel Science Foundation, Murray B. Koffler Professorial Chair.\\

\noindent
{\bf Acknowledgements.} 
This paper would be impossible without my students (in alphabetical order)
Yael Avni,
Nimrod Nir,
Itay Griniasty,
Sahar Sahebdivan,
and
William Simpson.
I also thank Mikhail Isachenkov, Ephraim Shahmoon, and Anna and Yana Zilberg for discussions, support and inspiration. 


\end{document}